\def\insertplot#1#2#3#4#5#6#7{
\vskip 10pt\nobreak\hbox to \hsize{\hss\dimen0=#3in\hbox to #6\dimen0{%
\dimen0=#2in\vbox to #6\dimen0{\vss
\special{ps: plotfile #1}
\special{ps::[end]
  PGPLOT restore
}
}\hss}\hss}\vskip 10pt}

\documentclass[runningheads]{cl2emult}

\usepackage{makeidx}  
\usepackage{graphicx} 
\usepackage{subeqnar} 
\usepackage{multicol} 
\usepackage{cropmark} 
\usepackage{lnp}      
\makeindex            

\begin{document}
\title*{Structure and Evolution of Circumstellar Disks Around Young Stars:
        New Views from ISO
\thanks{ISO is an ESA
project with instruments funded by ESA Member States (especially the PI
countries: France, Germany, the Netherlands and the United Kingdom) and
with the participation of ISAS and NASA.}}

\toctitle{Structure and Evolution of Circumstellar
\protect\newline Disks Around Young Stars}
%
%
\titlerunning{Structure and Evolution of Circumstellar Disks}
%
\author{Michael R. Meyer\inst{1} 
\thanks{Hubble Fellow} 
\and Steven V.W. Beckwith\inst{2}}
\authorrunning{Meyer et al.}
%
%
\institute{Steward Observatory, The University of Arizona, Tucson, AZ 85721--0065 USA \\
\and Space Telescope Science Institute, Baltimore, MD 21218 USA }

\maketitle              

\begin{abstract}
A question central to understanding the origin of our solar system
is: how do planets form in circumstellar disks around young stars?  
Because of the complex nature of the physical processes involved, 
multi--wavelength observations of large samples will be required
in order to obtain a complete answer to this question. 
Surveys undertaken with ISO have helped to solve pieces of this puzzle 
in addition to uncovering new mysteries.  We review a variety of studies 
aimed at understanding; i) the physical structure and composition of 
circumstellar disks commonly found surrounding young stellar objects; and ii)
the evolution of circumstellar disks from the active accretion 
phase to post--planet building debris disks. 
\end{abstract}

\section{Introduction}

The number of researchers involved in studies of circumstellar disks 
surrounding young stars has exploded in recent years -- and with good reason!
Discoveries of giant planets orbiting main sequence stars in the solar
neighborhood, of planetary mass companions surrounding pulsars, and 
of brown dwarf companions to stellar mass objects have brought new 
urgency and interest in the search to understand the structure and evolution 
of circumstellar disks.  In addition to their fundamental importance as
the likely sites of planet formation, circumstellar
accretion disks also play an important role in pre--main sequence evolution.
Depending on their mass, accretion rates, and lifetimes, such disks could
contribute significantly to building up the final mass of a solar type 
star.  They also appear to play a crucial role in regulating stellar angular 
momentum during the early accretion phase.  ISO, the first 
multi--mode infrared space observatory, devoted a significant 
fraction of time to object--oriented 
surveys of stars in order to study their circumstellar disks. 

We begin with a review of what was known and unknown about 
circumstellar disks surrounding pre--main sequence stars when 
ISO was launched.  In section 3, we discuss new results from 
ISO that shed light on the physical structure and composition of 
these disks.  In section 4, we review several mid-- and far--infrared 
surveys conducted with ISO focussed on the temporal evolution of the disks.  
Finally in section 5, we summarize these results and 
comment on progress that will be made in the coming years.   

\section{Properties of Circumstellar Disks}

It is generally accepted that the formation of a circumstellar disk
is a common outcome of the star formation process \cite{1996Natur.383..139B}.
Observational evidence for the existence of 
these disks abounds.  Many young stars exhibit infrared and millimeter 
excess emission indicating the presence of large amounts of circumstellar 
material.   Yet in order to provide an optically--thin 
line of sight toward the central object, a flattened dust distribution
is required \cite{1987ApJ...319..340M}.  Emission--line profiles observed 
toward young stars provided additional evidence.  Red--shifted components
of the bi--polar flow are thought to be occulted by a geometrically--thin, 
optically--thick disk resulting in blue--shifted line profiles 
\cite{1984A&A...141..108A}.  In young stellar objects where 
the observed flux is dominated by accretion luminosity
(e.g. FU Ori eruptive variables) 
one can observe kinematic signatures of rotation in 
absorption lines arising in the disk photosphere \cite{1990ApJ...349..328W}. 
However the most compelling evidence comes from direct images of the 
disks themselves (Figure 1). 

\begin{figure}
\centering
\insertplot{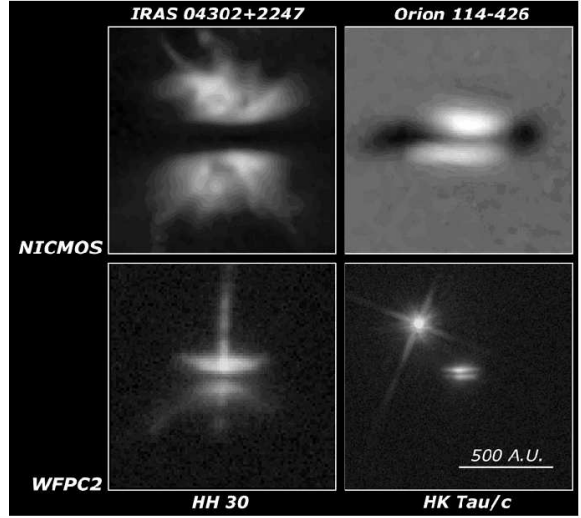}{5.0}{7.}{-2.50}{-0.4}{0.5}{1}
\caption[]{Optical and infrared images of circumstellar disks 
seen via scattered light with the Hubble Space Telescope: 
Images courtesy of NASA, STScI/AURA, and the following authors 
\cite{1999AJ....117.1490P} (upper--left), 
\cite{1998ApJ...492L.157M} (upper--right), 
\cite{1996ApJ...473..437B} (lower--left), and
\cite{1998ApJ...502L..65S} (lower--right).} 
\label{hstdisks}
\end{figure}

Based on these and other studies over the past 15 years, we have learned
a great deal concerning the physical properties of disks.  From sub--mm 
observations of optically--thin dust emission, estimates of total disk mass
(gas+dust) range from 0.1--10$^{-3}$ M$_{sun}$ (\cite{1995ApJ...439..288O}; 
\cite{1994ApJ...420..837A}; and \cite{1996A&A...309..493D}). 
Based on direct images obtained with HST, modeling of spectral 
energy distributions, and millimeter wave observations 
of gas disks, sizes are estimated to range from 10 $ < r < $ 1000 AU. 
Finally, observations of UV/blue excess emission allow one to 
estimate accretion rates from the disk onto the star 
\cite{1998ApJ...492..323G}.  
Recent estimates range from 10$^{-9}$ to 10$^{-7}$ M$_{sun}$ yr$^{-1}$ for 
classical T Tauri stars and $\times$ 100 greater for the FU Ori objects.  
There is a hint of decreasing accretion rate with age for T Tauri stars 
found in the Taurus--Auriga and Chamaeleon I complexes 
\cite{1998ApJ...495..385H}. 
This is consistent with the pioneering studies of the frequency 
of near--infrared excess emission 
\cite{1989AJ.....97.1451S} as a function 
of stellar age.  Near--IR excess emission (1--5 $\mu$m) traces 
dust emission at radii $<$ 0.1 AU and there is a
nearly 1:1 correlation between the presence of dust in the inner 
disk and spectroscopic signatures of accretion 
\cite{1995ApJ...452..736H}.  Recent studies which utilize larger statistically 
significant samples confirm that the timescale for dissipation 
of inner accretion disks is $<$ 30 Myr \cite{1999AAS...195.0209H}. 
Further, due to the small number of objects found
to be evolving from optically--thick to optically--thin 
disks in the region 0.1--1.0 AU, it appears that the 
transition time is $<$ 10$^6$ yrs 
\cite{1990AJ.....99.1187S}. 
Finally, there appears to be a connection between the 
presence/absence of an inner accretion disk and the evolution of stellar 
angular momentum \cite{1993AJ....106..372E}.   This relationship can be 
understood in terms of a magnetospheric star--disk interaction 
\cite{2000PPIV...JN}. 

Despite the tremendous explosion in our knowledge concerning 
circumstellar disks surrounding young stars, many fundamental 
questions remain unanswered.  What physical processes control 
the energy budget as a function of radius in the disk? 
What is the chemical composition of the dust grains observed?  
Does the evolution of disks in the planet--forming regions from 
0.1--10.0 AU differ from the evolution of the inner 
accretion disks?  ISO has made fundamental contributions toward 
answering these questions as described below. 

\section{Physical Structure and Composition} 

Photometric observations obtained 
over a broad wavelength range can be a powerful diagnostic 
of the spatial distribution of circumstellar material.  
As the temperature of the circumstellar material decreases with 
radius, progressively longer wavelength emission traces larger 
radii in the disk (Figure 2).  Utilizing the ISOPHOT instrument on--board ISO 
\cite{1996A&A...315L..64L}, Beckwith and collaborators set out 
to conduct a small survey of young star+disk systems 
in order to search for structure in the SEDs impossible to observe 
from the ground. They obtained multi--wavelength 
photometry with 13 filters from 4.9--160 $\mu$m for a sample of 14 objects
in the Taurus--Auriga and Chamaeleon I dark clouds (d $\sim$ 150 pc). 
Preliminary results are presented in Figure 3 for a representative 
sample of objects compared to expected stellar SEDs as well as 
a simple model of a geometrically--thin, optically--thick passive 
reprocessing disk seen face--on \cite{2000...MR}.  
A few general trends are 
obvious from inspection of these data.  First of all, several objects lack significant 
near--infrared excess emission suggesting the presence of inner
holes in the circumstellar dust distribution 
\cite{1997AJ....114..288M}. 
An extreme example is CS Cha in which 
the disk appears to be evacuated to $>$ 0.1 AU.  Secondly, it appears that 
a standard reprocessing (or accretion disk) matches the SEDs between 
2--10 $\mu$m.   Finally, the observed SEDs are flatter and 
exhibit greater flux in the mid-- and far--IR than simply blackbody disk 
models predict; an additional emission component is required.  
Similar conclusions have been reached from multi--wavelength study of u
the peculiar young star UX Ori \cite{1999A&A...350..541N} 
utilizing ISOPHOT. 

\begin{figure}
\centering
\insertplot{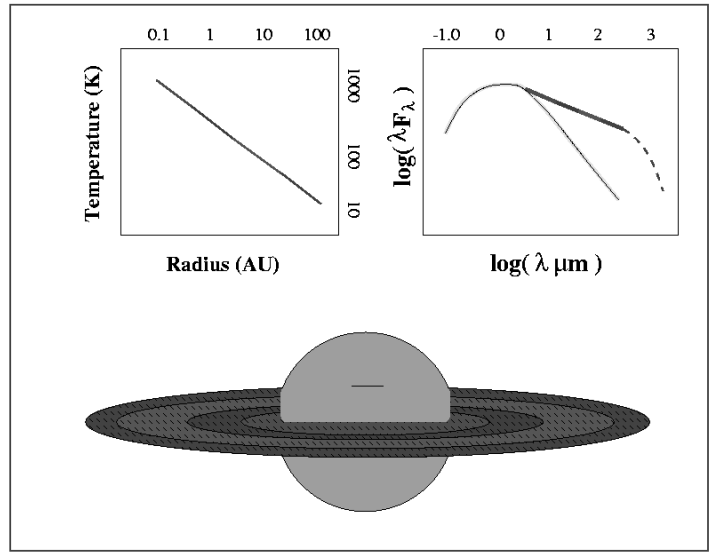}{8.0}{10.}{-0.0}{-0.3}{0.4}{1}
\caption[]{Schematic representation of an optically--thick, geometrically--thin 
circumstellar disk model.  Viscous accretion and/or reprocessing of stellar 
photons results in a power--law temperature distribution in the disk 
\cite{1987ApJ...312..788A}.  Each radius corresponds to a unique blackbody 
temperature which is diagnosed using a specific range of wavelengths
(see \cite{1999...SVWB}). 
}
\label{temp_rad}
\end{figure}

\begin{figure}
\centering
\insertplot{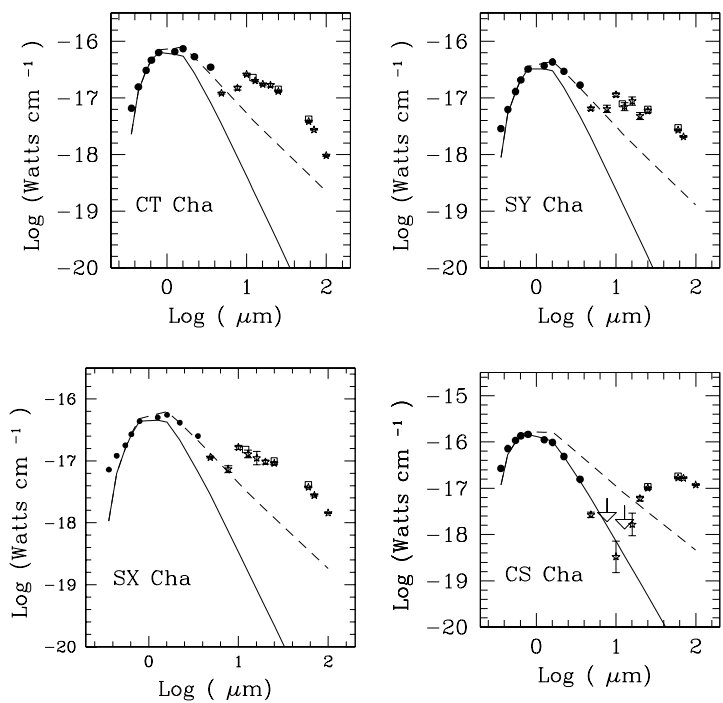}{7.0}{10.}{-0.20}{2.6}{0.5}{0}
\caption[]{Dereddened spectral energy distributions where the circles 
are ground--based photometry, the stars are new ISO photometry, and 
the squares are IRAS data.  Also shown are expected stellar photospheric 
flux appropriate for the known spectral type of the star (solid line) 
as well as expected SEDs from a geometrically--thin, optically--thick 
reprocessing disk viewed face--on (dashed line). 
}
\label{ltseds}
\end{figure}

What controls the energy budget in these circumstellar disks? 
For a disk  dominated by the dissipation of accretion energy confined to the 
disk mid--plane, the disk should exhibit a decreasing
temperature distribution away from the mid--plane which could produce absorption features
in the disk photosphere as observed in the FU Ori objects. 
For a disk dominated by reprocessing of stellar radiation, there should be 
a temperature inversion producing emission--lines in a hot, optically--thin 
disk atmosphere (\cite{1992RMxAA..24...27C}; \cite{1997ApJ...490..368C} (CG97)). 
A pioneering mid--IR spectral survey of T Tauri star+disk systems 
\cite{1985ApJ...294..345C}  found several sources exhibiting 
10 $\mu$m emission attributed to Si--O stretching modes.  
In order to investigate inner disk regions as well as understand 
the heating mechanisms which control disk structure, Natta, Meyer, and 
Beckwith \cite{2000ApJ...AN} utilized the PHOT--S module of ISOPHOT, to conduct a low 
resolution spectrophotometric survey from 2.5--11.7 $\mu$m of a sub--set 
of the T Tauri sample sample described above (see also \cite{1999A&A...346..205G}). 
{\it Each star in the sample of nine
exhibited some evidence of a 10 $\mu$m emission feature}.  
Using the simple flared--disk atmosphere model of CG97 and adopting a 
fiducial Si--O feature cross--section ($\sigma_\nu / \sigma_{10}$), 
they fitted these data for the 
efficiency of converting stellar photons into silicate emission 
($\epsilon = \sigma_{10} / \sigma_*$).   The results of these fits
are shown in Figure 4 where the 10 $\mu$m features are compared 
with the models.  The CG97 models successfully reproduce the majority of the observed 
spectra using a mixture of amorphous olivine and pyroxene grains 
with sizes $<$ 1 $\mu$m.  Combining these results with the 
SEDs described above suggests that disk atmospheres contribute 
significantly to the mid-- and far--infrared fluxes observed 
in young star+disk systems. 

\begin{figure}
\centering
\insertplot{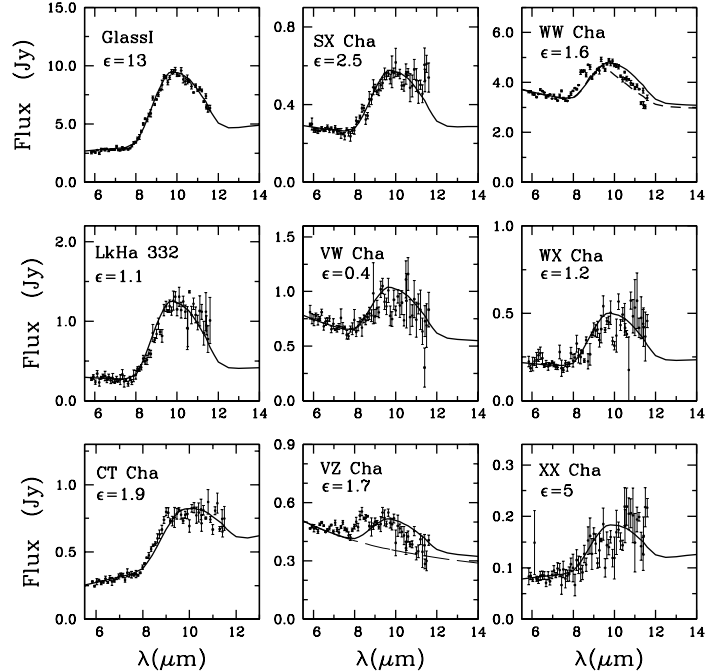}{6.0}{9.}{-0.5}{2.7}{0.5}{0}
\caption[]{
Model fits to the observed spectra. 
In each panel, the dots show the observed points 
while the solid line shows the predicted emission of the superheated
disk atmosphere, computed following CG97 for
pyroxene grains of 1.2 $\mu$m radius. 
The  value of $\epsilon =\sigma_{10}/\sigma_*$ which gives the best
fit to the observations is indicated. 
}
\label{models}
\end{figure}

Additional surveys undertaken with the SWS instrument on--board
ISO have enabled us to ask new questions regarding the dust 
mineralogy and gas content of these disks.  Moderate resolution
($R \sim 1500$) mid--infrared spectroscopy obtained for a sample
of young intermediate mass Herbig Ae/Be stars, 
have revealed the presence of crystalline silicate features
in three objects;  HD 100546, HD 142527, and HD 179218 
(\cite{1998A&A...332L..25M}; \cite{1999A&A...345..181M}; 
and \cite{1998Ap&SS.255...25W}). 
Similar features have been observed in the disk of 
$\beta$ Pictoris \cite{1993ApJ...418..440K} as well as in solar system comets
\cite{1994ApJ...425..274H}.  These observations are intriguing because 
crystalline silicates are not observed in the diffuse interstellar
medium.  Thus their presence in the circumstellar material surrounding 
the Herbig stars suggests that amorphous silicate
grains were transported into the inner disk region where they could be 
processed at temperatures $>$ 1500 K.  This, along with 
their observation at characteristic dust temperatures of $\sim$ 300 K, 
suggests large--scale radial mixing in these disks
(however see \cite{1999Natur.401..563M}).   Finally, we note that pioneering observations
have detected H$_2$ toward some young star+disk systems 
(\cite{1999ApJ...521L..63T}; \cite{1998Ap&SS.255...77V}). 
The relative intensities of the S(0) and S(1) lines suggest gas temperatures of
$\sim$ 100 K and mass estimates $\sim$ 0.01 M$_{sun}$ comparable to the
minimum mass solar nebula \cite{1985prpl.conf.1100H}. 

\section{Temporal Evolution} 

While it appears that most young stars
form accompanied by a circumstellar disk, it 
is not clear that all disks form planets.  
Several factors could play a crucial role in 
determining the fate of an accretion disk.  
It has been suggested that disks evolve more
quickly around high mass stars (M$_* >$ 1.0 M$_{sun}$)
compared to low mass stars \cite{2000PPIV...AN}.  Stellar companions 
can also effect the dynamical evolution of disk material 
\cite{2000PPIV...RDM}.  Finally, stellar environment 
can also play a role \cite{1998AJ....116.1816H}.
Of course one would like to 
study disk evolution as a function of all these variables.
Here we review the observed correlations of disk properties
with time as studied from ISO surveys of stars in clusters
with ages determined from evolutionary models as well as 
surveys of main sequence field stars of uncertain age. 

Despite significant advances in our understanding of 
the evolution of inner accretion disks {\it very little 
is known concerning the evolution of outer disks}.  At mid--infrared 
through sub--mm wavelengths (10--1000 $\mu$m) observations trace 
material between 0.1--10 AU.  Several surveys have been conducted 
with ISOPHOT which have made significant contributions to 
our understanding of outer disks.  Meyer et al.  \cite{2000...MRM} 
observed between 10 and 30 stars in each of five separate 
clusters and associations at 25 and 60 $\mu$m tracing material 
in the terrestrial planet zone from 0.3--3.0 AU.
This survey provides a complete census of optically--thick 
circumstellar disks at these wavelengths surrounding solar--type
stars with ages 1--300 Myr.  Despite being a factor of $\times$ 5
more sensitive than IRAS at 60 $\mu$m, only four sources were 
detected with $SNR > 3$ out of a sample of 97 stars.   
All of the sources detected were members of the Chamaeleon I dark 
cloud ($\tau < 10$ Myr) including one transition object 
possessing a large hole in its inner disk.  
These results suggest that optically--thick 
outer disks dissipate, or coagulate into larger particles on 
timescales comparable to the cessation of accretion (c.f. 
\cite{2000PPIV...SVWB}; \cite{2000PPIV...DH}). 
A complementary survey undertaken by Spangler et al. 
\cite{1999ESA...CS}, observed 
300 stars in young clusters and the field at 60 and 90 $\mu$m spanning 
a similar age range.  These authors analyze the data in terms of the 
fractional contribution of the far--infrared emission to the bolometric 
luminosity of the sources
observed ($f = L_{FIR}/L_{bol}$).  They plot the mean value of $f$ as a 
function of cluster age, and find a smooth variation from the youngest
stars ($f > 10^{-2}$ at ages $< 10^7$ yrs) to the oldest stars in 
sample ($f \sim 10^{-5}$ at ages $\sim 10^9$ yrs).  Presented in this
way, the zodiacal dust disk (within 5 AU) would have a value of 
$f \sim 10^{-7}$ for the 4.5 Gyr age of our solar system.  

\begin{figure}
\centering
\insertplot{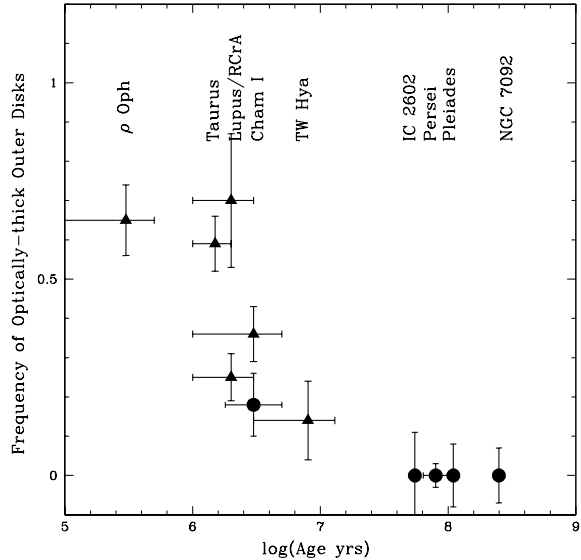}{7}{9}{0.2}{1.8}{0.4}{0}
\caption[]{
Frequency of 60 $\mu$m detections as a function of cluster
age for several star--forming regions observed by IRAS (triangles)
and ISO (circles) \cite{2000...MRM}.  Because the ISO observations
are $\times$ 5 more sensitive than the IRAS survey, the triangles should
be considered lower--limits when compared to the circles.  It appears 
that optically--thick disks from 0.3--3.0 dissipate or coagulate into
larger bodies on a timescale comparable to the termination of the main 
accretion phase in T Tauri disks.}
\label{outer}
\end{figure}

Other relevant surveys include the study of intermediate
mass members of the Ursa Major Stream ($\tau \sim 300$ Myrs) 
\cite{1998A&A...338...91A}.  One out of nine stars were detected consistent with 
a recent estimate of the frequency of IR excess among field main 
sequence stars \cite{1999A&A...343..496P}.  Similar results based on a survey of 
38 main sequence stars at 20 $\mu$m, tracing the inner--most annuli of these cool 
outer disks, have also been reported \cite{1999ApJ...520..215F}.  Finally, Habing et al. 
\cite{1999Natur.401..456H} present preliminary results concerning their survey of stars in 
the solar neighborhood.   Improving on searches for ``Vega phenomenon'' 
objects in the IRAS database, this survey was sensitive to stellar 
photospheric emission for A--K stars out to a distance
of 25 parsecs.  Combining ISO observations with improved age estimates for
their sample, they report the intriguing result that all stars younger 
than 300 Myr possess a detectable disk while no stars older than 400 Myr 
show IR excess emission at 60 $\mu$m.  They interpret this result as 
evidence that the era of maximum bombardment documented in our own 
solar system is a common phenomena in the evolution of circumstellar disks. 

\section{Summary and Future Work} 

From this vast array of observational results, a coherent picture
is emerging.  Most young stars (50--100 \%) possess active 
accretion disks during a significant portion of their PMS evolution.
These disks appear to have masses comparable to the minimum mass 
solar nebula in both gas and dust.  Structurally, they posses 
small inner holes ($<$ 0.05 AU) and flare in their outer regions
as expected from hydrostatic equilibrium.  Although almost all 
disks that extend to within 0.1 AU of the stellar surface appear
to be actively accreting, the energy budget of the outer regions
are dominated by reprocessing of light from the central star. 
The active accretion phase lasts from 1--10 Myr in most systems, 
though it can persist longer.  The time to transition from 
optically--thick to optically--thin within 1 AU lasts $<$ 1 Myr.  
At the same time, outer disks appear to dissipate, or grow into 
larger bodies (with decreased mass opacity) rendering them 
optically--thin.  The results of Meyer et al. \cite{2000...MRM} and 
Spangler et al. \cite{1999ESA...CS} appear to be consistent when comparing 
the frequency of detections in both surveys.  Additional 
data are required in the crucial 10--30 Myr old range order 
to discern whether or not there is a gradual evolution of
dust mass in small particles ($a <$ 1 mm) with time, or an 
abrupt change in disk properties associated with the termination
of accretion.  Perhaps particle growth in circumstellar disks 
is inhibited during the active accretion phase?  Similarly, 
further study of stars aged 100--500 Myr is necessary to follow--up 
the intriguing results of Habing et al. \cite{1999Natur.401..456H}.  Is the era of 
maximum bombardment, thought to be associated with 
with planet formation, a common phase of dust 
disk evolution around solar--type stars? 

Tremendous observational capabilities, both ground-- and space--based, 
will focus on these problems in the coming decade.  
With continued improvement in instrumentation, the current generation 
of 6--10m class telescopes will continue to make astounding discoveries 
such as the dramatic images of the disk surrounding HR4796A 
(\cite{1998ApJ...503L..83K}; \cite{1998ApJ...503L..79J}).  
High resolution images obtained from the UV through the 
near--infrared with HST will have an on--going 
impact on studies of circumstellar disks 
\cite{1999ApJ...525L..53W}.  Sub--millimeter telescopes (both single dish surveys 
and targeted observations with extant and future interferometric arrays) 
will provide data necessary to obtain a complete understanding 
of the evolution of dust mass and grain size in circumstellar disks 
\cite{1998Natur.392..788H}.  SIRTF
will enable a detailed census of faint debris disks in the solar 
neighborhood, as well as surveys of unprecedented sensitivity in 
nearby star--forming regions.  SOFIA will prove very complementary 
to SIRTF, providing higher spatial and spectral resolution.  
Finally, both ALMA and NGST 
have as central to their science missions 
studies of circumstellar disk evolution and planet formation.  
With the renewed interest in the origins of stars and planets, 
future work is sure to provide answers to questions that surveys 
undertaken with ISO have played a key role in defining. 

We would like to express our thanks to A. Natta, 
M. Robberto, B. Schulz, J. Stauffer, and D. Backman for their
continued collaboration in the analysis and interpretation of 
ISO data, and to the conference organizers for inviting this contribution. 
Special thanks to L.A. Hillenbrand for comments on an earlier 
version of this manuscript.  Support for this work was provided
by NASA through Hubble Fellowship Grant \# HF--01098.01--97A
awarded by the Space Telescope Science Institute which is operated
by the AURA, Inc., for NASA under contract NAS 5--26555.

\clearpage
\addcontentsline{toc}{section}{Index}
\flushbottom
\printindex

\end{document}